\documentclass[12pt]{article}
\usepackage{amssymb,amsmath}
\begin{document}

%%%%%%%%%%%%%%%%%%%%%%%%%%%%%%%%%%%%%%%%%%%%%%%%%%%%%%%%%%%%%%%%%%%%%
%%%%%%%%%%%%%%      TITLE PAGE     %%%%%%%%%%%%%%%%%%%%%%%%%%%%%%%%%%
%%%%%%%%%%%%%%%%%%%%%%%%%%%%%%%%%%%%%%%%%%%%%%%%%%%%%%%%%%%%%%%%%%%%%

\sloppy
\title
%{\hfill{\normalsize\sf FIAN/TD/01-15}    \\
 %           \vspace{1cm}
{\Large  On the problem of uniqueness of energy-momentum tensor of
 gravitational field }

\author
 {
       A.I.Nikishov
          \thanks
             {E-mail: nikishov@lpi.ru}
  \\
               {\small \phantom{uuu}}
  \\
           {\it {\small} I.E.Tamm Department of Theoretical Physics,}
  \\
               {\it {\small} P.N.Lebedev Physical Institute, Moscow, Russia}
  \\
  %       {\it {\small} 117924, Leninsky Prospect 53, Moscow, Russia.}
 }
%
%--------------------------------------------------------------------
\maketitle
%--------------------------------------------------------------------
%%%%%%%%%%%%%%%%%%%%%%%%%%%%%%%%%%%%%%%%%%%%%%%%%%%%%%
\begin{abstract}
For an island-like distribution of matter the gravitational energy-momentum 
tensor is  defined according to Weinberg as a source of 
metric. If this source is formed by self-interactions of gravitons, so that
nonphysical degrees of freedom are excluded, then this source is a reasonable
candidate for the energy-momentum tensor of gravitational field. 
The disastrous influence of the nonphysical degrees of freedom is demonstrated
by comparing the gravitational energy-momentum tensors in the harmonic,
isotropic and standard frames for the Schwarzschild solution. The harmonic 
frame is clearly preferable for defining the gravitational energy-momentum
tensor.
\end{abstract}

\section{The gravitational energy-momentum tensor as the source of metric}
There are several arguments in favor of non-localizability of the energy of
the gravitational field, see \S 20.4 in [1]. They do not seem convincing 
enough.

 Following [2], we consider the case when 
 $$
 g_{\mu\nu}=\eta_{\mu\nu}+h_{\mu\nu},\quad \eta_{\mu\nu}={\rm diag(-1,1,1,1)}
 \eqno(1)
 $$
 and $h_{\mu\nu}\to 0$ quickly enough when $x\to\infty$, but it is not assumed
 that $h_{\mu\nu}\ll1 $ everywhere.
 The wave equation for $h_{\mu\nu}$ is
 $$
 h_{\mu\nu,\lambda}{}^{\lambda}-h^{\lambda}{}_{\mu,\lambda\nu}-
 h^{\lambda}{}_{\nu,\lambda\mu}+h_{,\mu\nu}+
 \eta_{\mu\nu}(h_{\sigma\lambda}{}^{,\sigma\lambda}-h_{,\lambda}{}^{\lambda})=
 $$
 $$
 -16\pi G(T_{\mu\nu}+t_{\mu\nu}),\quad h\equiv h_{\lambda}{}^{\lambda};
 \quad h_{,\sigma}\equiv\frac{\partial}{\partial x^{\sigma}},\eqno(2)
 $$
 cf. Ch.3, \S 17 in [3].
 Here $t_{\mu\nu}$ is the gravitational energy-momentum tensor. 

 In general relativity eq. (2) is the Einstein equation with
 $$
 8\pi Gt_{\mu\nu}=R_{\mu\nu}-\frac12g_{\mu\nu}R-R_{\mu\nu}^{(1)}+\frac12
 \eta_{\mu\nu}R^{(1)\lambda}{}_{\lambda},                           \eqno(3)
 $$
 see equations (7.6.3) and (7.6.4) in [2].
 Indices in  $h_{\mu\nu}$, $R_{\mu\nu}^{(1)}$ and 
 $\frac{\partial}{\partial x^{\sigma}}$ are raised and lowered with the 
 help of $\eta$, while  indices of true tensors, such as $R_{\mu\nu}$ are 
 raised and lowered with the help of $g$ as usual. $R_{\mu\nu}^{(1)}$ is a
  linear in $h_{\mu\nu}$ part of $R_{\mu\nu}$:
  $$
  R_{\mu\nu}^{(1)}=\frac12[h_{,\mu\nu}-h^{\lambda}{}_{\mu,\lambda\nu}-
  h^{\lambda}{}_{\nu,\lambda\mu}+h_{\mu\nu,\lambda}{}^{\lambda}],\quad
 h=\eta^{\mu\nu}h_{\mu\nu}\quad,
 \quad R^{(1)}=\eta^{\mu\nu}R^{(1)}_{\mu\nu}.               \eqno(4)
  $$
 As shown in [2], tensor (3) has all the necessary properties of a 
 gravitational 
 energy- momentum tensor. The same is true even if the general relativity
 is not assumed i.e. $t_{\mu\nu}$ has not the form (3). Non the less,
 $t_{\mu\nu}$ as defined in (3) has one drawback: it 
 depends on a coordinate system in an inadmissible way.
 I shall demonstrate this for the gravitational field of a spherically
 symmetric body and then indicate how, in my opinion, to correct the situation.

 To begin with, we write down some of metrics outside the body. \\
 1. Standard Schwarzschild
 $$  
 d\tau^2=(1-\frac{2GM}{r})dt^2-r^2(\sin^2\theta d\varphi^2+d\theta^2)-
  \frac{dr^2}{1-\frac{2GM}r}.                                         \eqno(5)
 $$
 2. Harmonic ($R=r-GM$)
 $$
 d\tau^2=\frac{1-\frac{GM}R}{1+\frac{GM}{R}}dt^2-
 \left(1+\frac{GM}{R}\right)^2R^2
 (\sin^2\theta d\varphi^2+d\theta^2)-\frac{1+\frac{GM}{R}}{1-\frac{GM}R}dR^2.
                                                                       \eqno(6)
 $$
 3. Isotropic ($r=\rho(1+\frac{GM}{\rho})^2$)
 $$
 d\tau^2=\left(\frac{1-\frac{GM}{2\rho}}{1+\frac{GM}{2\rho}}\right)^2dt^2-
 \left(1+\frac{GM}{2\rho}\right)^4[d\rho^2+\rho^2
 (\sin^2\theta d\varphi^2+d\theta^2)].                              \eqno(7)
 $$
 4. Eddington frame
 $$
 d\tau^2=dt^{*2}-\frac{2GM}{r^*}(dr^*+dt^*)^2-dr^{*2}-r^{*2}d\Omega.\eqno(8)
 $$
 It is remarkable that in this frame $g_{\mu\nu}$ linearly depends upon G.
 Although all frames are equal, at least for our purpose there is a
  frame that is "more equal than others".

 Now we go back to equation (2) and consider space region where $T_{\mu\nu}=0$. 
 Then the equation tell us that if we know $h_{\mu\nu}$, we can obtain 
 $t_{\mu\nu}$. Using this, we get the 
 gravitational energy-momentum tensors in 
 the harmonic, isotropic and standard Schwarzschild frames. Instead of 
 spherical coordinates in (5-7) we use below the rectangular ones and
 denote the space coordinates by $x_i,\quad x_ix_i=r^2, \quad i=1, 2, 3$ 
  {\it in all three frames}.

 \section{The harmonic frame}

 In this frame the gravitational energy-momentum tensor was obtained in [4].
 We have
 $$
 h_{00}=\frac{-2\phi}{1-\phi},\quad h_{ij}=(-2\phi+\phi^2)\delta_{ij}+
 \frac{1-\phi}{1+\phi}\phi^2\frac{x_ix_j}{r^2},\quad
 h_{0i}=0,\quad \phi=-\frac{GM}{r}.
                                                                \eqno(9)
 $$
 $$
 8\pi Gt_{00}^{har}=\frac{1}{r^2}[-2-\phi^2+\frac4{1+\phi}-\frac2{(1+\phi)^2}],
 \quad \quad t_{0i}^{har}=0;                       \eqno(10)
 $$
 $$
 8\pi Gt_{ij}^{har}=\frac{x_ix_j}{r^4}\left(-2\phi^2+2-\frac1{1+\phi}-
 \frac1{(1+\phi)^2}+\frac1{1-\phi}+\frac1{(1-\phi)^2}-
 \frac2{(1-\phi)^3}\right)+
 $$
 $$
 \frac{\delta_{ij}}{r^2}\left(\phi^2-\frac1{1+\phi}+
 \frac1{(1+\phi)^2}+\frac1{1-\phi}-\frac3{(1-\phi)^2}+
 \frac2{(1-\phi)^3}\right),\quad i,j=1,2,3.                      \eqno(11)
 $$
 We see that the energy-momentum tensor has singularities at $r=GM$, when 
 $\phi=-1$.

 For $\phi\ll 1$ we get
 $$
 \left.8\pi Gr^2t_{00}^{har}\right|_{\phi\ll1}=-3\phi^2+4\phi^3+\cdots; 
                                                    \eqno(12)
 $$
  $$
  \left.8\pi Gt_{ij}^{har}\right|_{\phi\ll1}=
  7\frac{\phi^2}{r^2}\left(\delta_{ij}
  -\frac{2x_ix_j}{r^2}\right)+6\frac{\phi^3}{r^2}\left(\delta_{ij}-
  \frac{5}{3}\frac{x_ix_j}{r^2}\right)+\cdots.                 \eqno(13)
  $$
 Defining $\tilde h^{\mu\nu}$ by
 $$
\tilde g^{\mu\nu}\equiv(-g)^{1/2}g^{\mu\nu}=\eta^{\mu\nu}+
\tilde h^{\mu\nu},\quad g={\rm det}g_{\mu\nu}, \eqno(14)
 $$
 we get the harmonic condition in the form
 $$
 \tilde h^{\mu\nu},_{\nu}=0.                 \eqno(15)
 $$

 This condition is the analogue of Lorentz condition in electrodynamics.
 It exclude the nonphysical degrees of freedom and the simplest assumption is
 that in our case it exclude all the nonphysical degrees. In this case
  the rectangular 
 harmonic frame is the preferred system, the use of which has been
  advocated by Fock [5]. 
 In general relativity we have
 $$
 \tilde g^{ik}=\delta_{ik}-\phi^2\frac{x_ix_k}{r^4},  \eqno(16)
 $$
 see \S 58 in [5]. This expression can be obtained also from the second
 order of perturbation expansion, considered in [6] using quantum tree graphs.

 If general relativity is not assumed, we have instead of (16)
 $$
 \tilde g^{ik}=\delta_{ik}-\alpha\phi^2\frac{x_ix_k}{r^4},  \eqno(17)
 $$
 where $\alpha$ is a coefficient of order one. It is determined by 
 a chosen 3 graviton vertex. The higher terms in the expansion (17),
 i.e. terms with $\phi^n$ and $\phi^n\frac{x_ix_k}{r^2}$, $n>2$ are still
 absent due to (15) as seen from the relations
 $$
 \left(\frac1{r^n}\right){}_{,j}=-\frac{nx_j}{r^{n+2}},\quad
 \left(\frac{x_ix_j}{r^n}\right){}_{,j}=\frac{x_i(4-n)}{r^{n}}.
 $$
 Only term with $n=4$,  i.e. term $\phi^2\frac{x_ix_k}{r^4}=G^2M^2x_ix_k/r^4$
 is possible.

 Using (17) and letting $g_{00}$ to be arbitrary for the time being,
 we can express $g_{ik}$ through it. First, from (17) we find
 $$
 {\rm det}\tilde g^{ik}=1-\alpha\phi^2. \eqno(18)
 $$
 Then we have
 $$
 {\rm det}\tilde g^{\mu\nu}=\tilde g^{00}{\rm det}\tilde g^{ik}=g. \eqno(19)
 $$

 Then, using $\tilde g^{00}=\sqrt{(-g)}g^{00}=\sqrt{-g}/g_{00}$,
  we obtain from (18) and (19)
 $$
 \sqrt{-g}=-\frac{1-\alpha\phi^2}{g_{00}}.
 $$
 From definition $\tilde g^{ik}\tilde g_{kj}=\delta_{ij}$ we find
 $$
 \tilde g_{kj}=\delta_{kj}+
 \frac{\alpha\phi^2}{1-\alpha\phi^2}\frac{x_kx_j}{r^2}.  \eqno(20)
 $$
 Finally, we have
 $$
 g_{ik}=\sqrt{-g}\tilde g_{ik}=-\frac1{g_{00}}[(1-\alpha\phi^2)\delta_{ik}+
 \alpha\phi^2\frac{x_ix_k}{r^3}].    \eqno(21)
 $$

 Assuming $g_{00}=-\exp{(2\phi)}$, obtained in [7] from heuristic
  considerations, we get $g_{\mu\nu}$ and $t_{\mu\nu}$ regular everywhere
  except $r=0$. The same form of $g_{00}$ appears in general relativity
  for a model of a spherical body considered in the cylindrical 
  coordinates, see eq. (8.30) in [8].

  In any case, from perihelion precession we know that in $G^2$ approximation
  $g_{00}=-(1+2\phi+2\phi^2)$. Then (21) in this approximation gives
  $$
  g_{ik}^{(2)}=[1-2\phi+(2-\alpha)\phi^2]+\alpha\phi^2\frac{x_ix_k}{r^2}.
                                                                    \eqno(22)
  $$
  In general relativity $\alpha=1$. If we want to preserve the coordinate
  condition (15) in a more general approach, $g_{ik}^{(2)}$  still must have
  the form (22).
   
 \section{Isotropic frame}

 In this frame 
 $$
 h_{00}=\frac{4}{1-\frac{\phi}2}-\frac{4}{(1-\frac{\phi}2)^2},\quad
 h_{ij}=\delta_{ij}[-2\phi+\frac32\phi^2-\frac12\phi^3+\frac1{16\phi^4}],\quad
 h_{0i}=0.
                                                               \eqno(23)
 $$  
 Here $\phi$ has the same form as in (9).
 From here we get
 $$
 h_{ij,km}=\delta_{ij}\frac{x_kx_m}{r^4}[-6\phi+12\phi^2-\frac{15}2\phi^3
 +\frac32\phi^4]+
 %$$
 %$$
 +\frac{\delta_{ij}\delta_{km}}{r^2}[2\phi-3\phi^2+\frac{3}2\phi^3
 -\frac14\phi^4],                 \eqno(24)
 $$
  $$
  h_{00,ij}=\frac{\delta_{ij}}{r^2}[\frac{4}{1-\frac{\phi}2}
  -\frac{12}{(1-\frac{\phi}2)^2}+\frac{8}{(1-\frac{\phi}2)^3}]+
  \frac{x_ix_j}{r^4}[-\frac{4}{1-\frac{\phi}2}
  -\frac{4}{(1-\frac{\phi}2)^2}+\frac{32}{(1-\frac{\phi}2)^3}
 - \frac{24}{(1-\frac{\phi}2)^4}].       \eqno(25)
  $$
 and 
 $$
 h=-\frac{4}{1-\frac{\phi}2}+\frac{4}{(1-\frac{\phi}2)^2}+
 3[-2\phi+\frac32\phi^2-\frac12\phi^3+\frac1{16}\phi^4],       \eqno(26)
 $$
 
 Using these expressions in (4), we find 
 $$
 R^{(1)}_{00}=\frac12h_{00,ii}=\frac1{r^2}[\frac{4}{1-\frac{\phi}2}
  -\frac{20}{(1-\frac{\phi}2)^2}+\frac{28}{(1-\frac{\phi}2)^3}-
 - \frac{12}{(1-\frac{\phi}2)^4}],                           \eqno(27) 
 $$
 $$
 R^{(1)}_{ij}= \frac{x_ix_j}{r^4}[\frac{2}{1-\frac{\phi}2}
  +\frac{2}{(1-\frac{\phi}2)^2}-\frac{16}{(1-\frac{\phi}2)^3}
 + \frac{12}{(1-\frac{\phi}2)^4}-3\phi+6\phi^2-\frac{15}4\phi^3
 +\frac34\phi^4]+ 
 $$
 $$
 \frac{\delta_{ij}}{r^2}[-\frac{2}{1-\frac{\phi}2}
  +\frac{6}{(1-\frac{\phi}2)^2}-\frac{4}{(1-\frac{\phi}2)^3}
 + \phi-\frac{3}4\phi^3
 +\frac14\phi^4]                                            \eqno(28)
 $$
 Finally, we obtain from (3) (the first two terms on the r.h.s. of (3)
 disappear because we use the exact solution of the Einstein equation
  and consider the region of space without matter)
 $$
 8\pi Gr^2t_{00}^{iso}=-3\phi^2+3\phi^3-\frac34\phi^4,\quad \phi=
 -\frac{GM}{r},
 \quad t_{0i}^{iso}=0;                             \eqno(29)
 $$
 (this is an exact expression for $t_{00}^{iso}$) and similarly 
$$
8\pi Gt_{ij}^{iso}
=\frac{x_ix_j}{r^4}[-\frac2{1-\frac{\phi}2}
-\frac2{(1-\frac{\phi}2)^2}+\frac{16}{(1-\frac{\phi}2)^3}-
\frac{12}{(1-\frac{\phi}2)^4}+3\phi-6\phi^2+\frac{15}4\phi^3-\frac34\phi^4]+
$$
 $$
 \frac{\delta_{ij}}{\rho^2}[-\frac2{1-\frac{\phi}2}
+\frac{14}{(1-\frac{\phi}2)^2}-\frac{24}{(1-\frac{\phi}2)^3}+
\frac{12}{(1-\frac{\phi}2)^4}-\phi+3\phi^2-\frac{9}4\phi^3+\frac12\phi^4].
                                                        \eqno(30)
$$

For $\phi\ll 1$ we have
$$
\left.8\pi Gt_{ij}^{iso}\right|_{\phi\ll1}=7\frac{\phi^2}{r^2}
\left(\delta_{ij}-2\frac{x_ix_j}{r^2}\right)+\frac{9}{2}\frac{\phi^3}{r^2}
\left(\delta_{ij}-\frac{5}{3}\frac{x_ix_j}{r^2}\right)  +\cdots     \eqno(31)
$$

 We note now that $t_{\mu\nu}^{iso}$ is regular everywhere, except at
$r=0$. This can be expected because the metric in (7) is regular.
The transformation from the harmonic frame
to the isotropic one is also regular transformation (as well near horizon): 
$R=\rho(1+\frac{GM}{2\rho})^2-GM$.
So  there is no reason for
 disappearance of singularities in $t_{\mu\nu}^{iso}$.
  If we assume that $h_{\mu\nu}^{har}$ is formed only by the physical 
degrees of freedom, we may consider $t_{\mu\nu}^{har}$ as a correct tensor and 
interpret the disappearance of singularity in $t_{\mu\nu}^{iso}$ as foul 
play of nonphysical degrees of freedom. 

From (12-13) and (29), (31) we  see that in $\phi^2$ approximation
 $t_{\mu\nu}^{har}$
coincides with $t_{\mu\nu}^{iso}$. This is in agreement with the fact that
 $h_{\mu\nu}^{(2)iso}$  (i.e.
 $h_{\mu\nu}^{iso}$ in the $\phi^2$ approximation) can be
 obtained from
$h_{\mu\nu}^{(2)har}$ by gauge transformation
$$
 h_{ij}^{(2)har}-h_{ij}^{(2)iso}=\frac12G^2M^2(\frac{\delta_{ij}}{r^2}
 -\frac{2x_ix_j}{r^4})=
 \frac14G^2M^2(\Lambda_{i,j}+\Lambda_{j,i}),\quad \Lambda_{i}=\frac{x_i}{r^2}.
                                                                  \eqno(32)
 $$
 Here 
 $$
 h_{ij}^{(2)har}=\phi^2\left(\delta_{ij}+\frac{x_ix_j}{r^2}  \right), \quad
    h_{ij}^{(2)iso}=\frac32\phi^2\delta_{ij},
 $$
 see (9) and (23).
 The gauge transformation does not changes the source and may be interpreted
  as a change of frame, but {\it not visa versa}. The linear approximation 
  in $h_{\mu\nu}$, i.e. $h_{\mu\nu}^{(1)}$ produces the $\phi^2$ approximation 
  in the source (see equation (7.6.15) in [2]) and that is
  why $t_{\mu\nu}^{(2)}$ coincide in both frames.

  \section{Standard frame}

 In this case we have
 $$
 h_{00}=-2\phi,\quad h_{ij}=\frac{x_ix_j}{r^2}\left(\frac1{1+2\phi}-1\right).
                                                               \eqno(33)
 $$
 Simple calculations give
 $$
 h_{ij,kl}=\frac{x_ix_jx_kx_l}{r^6}[-8+\frac{3}{1+2\phi}+\frac{3}{(1+2\phi)^2}
 +\frac{2}{(1+2\phi)^3}]+\frac{1}{r^4}(\delta_{il}x_jx_k+\delta_{jl}x_ix_k+
 $$
 $$
 \delta_{kl}x_ix_j+\delta_{ik}x_jx_l+\delta_{jk}x_ix_l)[2-\frac{1}{1+2\phi}
 -\frac{1}{(1+2\phi)^2}]
+\frac{1}{r^2}(\delta_{ik}\delta_{jl}+\delta_{jk}\delta_{il})[\frac{1}{1+2\phi}
-1].                              \eqno(34)
$$
As in previous Section we find
$$
8\pi Gt_{00}^{st}=\frac1{r^2}[-1+\frac{2}{1+2\phi}-\frac{1}{(1+2\phi)^2}],
                               \eqno(35)
$$
$$
\left.8\pi Gt_{00}^{st}\right|_{\phi\ll1}=\frac1{r^2}[-4\phi^2+16\phi^3+\cdots],
                                                              \eqno(36)
$$
and
$$
8\pi Gt_{ij}^{st}=\frac{x_ix_j}{r^4}[1-3\phi-\frac1{2(1+2\phi)}-
\frac1{2(1+2\phi)^2}]+\frac{\delta_{ij}}{r^2}[\phi-
\frac1{2(1+2\phi)}-\frac1{2(1+2\phi)^2}],  \eqno(37)
$$
$$
\left.8\pi Gt_{ij}^{st}\right|_{\phi\ll1}=
4\frac{\phi^2}{r^2}\left(\delta_{ij}-\frac{2x_ix_j}{r^2}\right)
-12\frac{\phi^3}{r^2}\left(\delta_{ij}-\frac{5}{3}\frac{x_ix_j}{r^2}\right)
 +\cdots].                                                  \eqno(38)
$$
Comparing (36) and (38) with corresponding expressions in harmonic frame,
see (12) and (13), we note that there is an essential difference even in
 $\phi^2$ approximation. This means that even 
 $h^{(1)st}$ cannot be obtained from $h^{(1)har}$ by gauge transformation.
  But the difference in radial coordinates in these
 systems [see the transition from (5) to (6)], i.e. $r-R=GM$, can't be
  responsible for the differences in
the  energy-momentum tensors when $R, r\gg GM$. The blame must be laid upon
   the violation of the coordinate condition (15), i.e. on the nonphysical
    degrees of freedom.

    Finally, we note that in the considered static field in space without 
    matter
    the conservation law $t^{\mu\nu}{}_{,\nu}$ takes the form $t_{ij,j}=0$
  outside the horizon.   Each term in the
     expansion of $t_{ij,j}$ in power series of $\phi$ must satisfy the
     conservation law and this dictates up to a constant factor  the form  
     of the $n-$th term of the expansion:
   $$
   \frac{\phi^n}{r^2}\left(\delta_{ij}-\frac{n+2}{n}\frac{x_ix_j}{r^2} \right). 
   $$
 \section{Conclusion}

  We see that the energy-momentum tensor of gravitational field requires the
  existence of privileged coordinate system and there are some grounds
   to assume that
  the rectangular harmonic frame is such a system. Of course, any other frame
  is also good if we deal with covariant quantities, but the 
  energy-momentum tensor must be properly transformed from the privileged
  system.  It seems reasonable    to expect that in 
  the region of applicability of any theory, its energy-momentum tensor 
  should be finite. More exactly we expect that the total gravitational
 energy in space outside a radius $r$ must be finite.
  In general relativity, with definitions of $t_{\mu\nu}$ which seems 
  reasonable, this energy goes to $-\infty$ when $r\to GM$ [9,4]. So 
  the consideration of possible deviations from general relativity are of
  interest, cf [10].
\section{Acknowledgments}
 This work was supported in part by the Russian Foundation for
Basic Research (projects no. 00-15-96566 and 02-02-16944).

 \section*{References}
\begin{enumerate}

\item  Misner C.W., Thorne K.S., Wheeler J.A., {\it Gravitation}, Sam Francisco
(1973). 
\item Weinberg S., {\it Gravitation and Cosmology}, New Yorc (1972).
\item Schwinger J., {\it Particle, Sources and Fields}, Addison-Wesley
Publishing Company, Reading (1970). 
\item Nikishov A.I. gr-qc/0002038; Bulletin of the Lebedev Physics Institute,
N12, pp. 45-49 (1999).
\item Fock V., {\it The Theory of Space-Time and Gravitation.}
(2-nd revised edition, Pergamon Press, New York, 1964.)
\item Duff M.J., Phys. Rev. {\bf D}, V. {\bf7}, 2317 (1972).  
\item Dehnen H., H\"onl H. and Westpfahl, Ann. der Phys. {\bf 6}, 7 Folge,
Band 6, Heft 7-8, S. 670 (1960).
\item Synge J.L., {\it Relativity: The General Theory}, Amsterdam (1960).
\item Dehnen H. Zeit. f\"ur Phys., {\bf179} Band, 1 Heft, S. 76 (1964).
\item Nikishov A.I., gr-qc/9912034; Part. and Nuclei, {\bf32},p. 5 (2001). 
\end{enumerate}
 \end{document}